\documentclass[10pt]{article}
\usepackage{amssymb}
\usepackage{amsfonts}
\usepackage{rotating}
 \usepackage{amsmath}
\usepackage{graphicx}
\usepackage{epsfig}
\usepackage{comment} 
\usepackage{graphicx}
\usepackage{cite}

\usepackage[labelfont=bf,labelsep=period,justification=raggedright]{caption}

\makeatletter
\renewcommand{\@biblabel}[1]{\quad#1.}
\makeatother

\date{}

\usepackage[usenames]{color}
\usepackage{url}

\newcommand{\be}{\begin{equation}}
\newcommand{\ee}{\end{equation}}

\newcommand{\bea}{\begin{eqnarray}}
\newcommand{\eea}{\end{eqnarray}}

\begin{document}

\begin{flushleft}
{\Large
\textbf{The Twitter of Babel: Mapping World Languages through Microblogging Platforms} 
}
\\
\vspace{0.2cm}
Delia Mocanu $^{1}$, 
Andrea Baronchelli$^{1}$, 
Nicola Perra$^{1}$,
Bruno Gon\c calves $^{2}$, 
Alessandro Vespignani$^{1,3,4}$
\\
\vspace{0.1cm}
\bf{1} Laboratory for the Modeling of Biological and Socio-technical Systems, Northeastern University, Boston, MA 02115 USA
\\
\bf{2} Aix Marseille Universit\'e, CNRS UMR 7332, CPT, 13288 Marseille, France
\\
\bf{3} Institute for Quantitative Social Sciences at Harvard University, Cambridge, MA 02138 USA
\\
\bf{4} Institute for Scientific Interchange Foundation, Turin, Italy
\\
\end{flushleft}

\begin{abstract} 
Large scale analysis and statistics of socio-technical systems that just a few short years ago would have required the use of consistent economic and human resources can nowadays be conveniently performed by mining the enormous amount of digital data produced by human activities. Although a characterization of several aspects of our societies is emerging from the data revolution, a number of questions concerning the reliability and the biases inherent to the big data ``proxies" of social life are still open. 
Here, we survey worldwide linguistic indicators and trends through the analysis of a large-scale dataset of microblogging posts. We show that available data allow for the study of language geography at scales ranging from country-level aggregation to specific city neighborhoods. The high resolution and coverage of the data allows us to investigate different indicators such as the linguistic homogeneity of different countries, the touristic seasonal patterns within countries and the geographical distribution of different languages in multilingual regions. This work highlights the potential of geolocalized studies of open data sources to improve current analysis and develop indicators for major social phenomena in specific communities.   
\end{abstract}


\section{Introduction}
Modern life, with its increasing reliance on digital technologies, is opening unanticipated opportunities for the study of human behavior and large scale societal trends. Cell phones have been playing a pivotal role in this revolution, serving as ubiquitous sensors, and  the default point of contact for online activities \cite{gonzalez08-1,Onnela:2007p4919}. As a whole, mobile clients for microblogging platforms, social networking tools, and other ``proxy" data of human activity collected in the web allow for the quantitative analysis of social systems at a scale that would have been unimaginable just a few years ago \cite{hale2012world, conover11-2, sang2012predicting,goncalves2011modeling}. In particular, the possibility of using mobile-enabled microblogging platforms, such as Twitter, as monitors of public opinion, social movements and as tools for the mapping of social communities has generated much interest in the literature \cite{borge2011structural,tumasjan2010predicting, culotta2010towards,salathe11-1,salathe12-2,kulshrestha2012geographic,mislove2011understanding,hong_2011}. At the same time it is crucial to understand to which extent the picture of socio-technical systems emerging from digital data proxies is a statistically sound and how well it does scale to a planetary dimension \cite{Giannotti2012}.

In this paper, we perform a comprehensive survey of the worldwide linguistic landscape as emerging from mining the Twitter microblogging platform. Our large-scale dataset,  gathered over approximately two years, at an average rate of $6.5 \times 10^5$ \textit{GPS-tagged} tweets per day, contains information about almost $6$ million users and provides a uniquely fine-grained survey of  worldwide linguistic trends. By coupling the geographical layer to the identification of the language of single tweets we are able to determine the detailed language geography of more than $100$ countries worldwide~\cite{williams88-1}. 

Although previous studies have investigated the language dynamics \cite{baronchelli2012language} of Twitter, those analysis have focused on specific, yet interesting, aspects concerning the combined study of language and geographical analysis in Twitter, and a global picture is still lacking. For instance, most represented languages have been identified for the Top-$10$ more active countries \cite{poblete2011all}, language-dependent differences have been pointed out in the user activity related to the posting and conversations patterns \cite{Weerkamp2011}, and language has been shown to be a strong predictor for the formation of follower/followee relations \cite{takhteyev2012geography}. 
For this reason and for the sake of assessing the generality and planetary scalability of our analysis, we have first focused on the reliability of geospatial trends extracted from our dataset. Interestingly, we find a universal pattern describing users' activity across countries, and a clear correlation between Twitter adoption and the Gross Domestic Product (GDP) of a country, further characterized by well defined continent-dependent trends. 

The high quality of the dataset permits the study of the spatial distribution of different languages at different scales from aggregated country-level analysis to the neighborhood scale. In particular we can drill down data of linguistic macro areas and single out heterogeneities at the country and regional level, scrutinizing the cases offered from Belgium and Catalonia (Spain) as examples. Furthermore we explore the resolution offered by the data at very fine level of granularity and inspect the city and neighborhood levels, taking as case studies the spatial distribution of French and English languages in Montreal (Canada) and inspecting linguistic majorities in New York City (USA). We find that Twitter is able to reproduce the geospatial adoption of languages for a wide range of resolution scales. We contrast our results  against census data, and discuss the possible sources of discrepancies between the two. Finally, we broaden our perspective by addressing the seasonality patterns in the language composition of the Twitter signal. We use  touristic countries such as Italy, Spain, and France to single out clear seasonal trends like, for instance, the increase of English and other languages during the summer holiday season. Overall, our  analysis highlights the potential of Twitter data in defining open source indicators for geospatial trends at the planetary scale.  

The paper is structured as follows. In section \ref{results} we go over data selection criteria as well as statistical measures regarding the universality of users behavior. Within this framework, we investigate several relevant examples in language geography (section \ref{sec:language_analysis}) and explore the temporal dimension for seasonal patterns (section \ref{sec:seasonal}). A discussion (section \ref{sec:discussion}) of the results is followed by a thorough description of the data sets and methodology used (section \ref{sec:methods}). 

\section{Results}
\label{results}

Our analysis is based upon Twitter data gathered in approximately $20$ months between October $18$, $2010$ and May $17$, $2012$, at an average rate of  $6.5 \times 10^5$ GPS-tagged tweets per day (see Table \ref{tab:geoloc} for exact numbers). The dataset includes $3.8 \times 10^8$ tweets produced by $6.0 \times 10^6$ users located in $191$ countries, $110$ of which generated the amount of data necessary for a significant statistical analysis of language detection. Our language detection methods allowed us to identify $78$ languages. Our analysis is restricted to GPS-tagged tweets in order to preserve maximum level of geographical detail, taking into account both live GPS updates and device stored locations. The amount of geolocalized signal could in fact be increased by considering different kinds of metadata, like for example self reported locations \cite{mislove2011understanding}, but these procedures would not allow us to reach the level of granularity and detail we aim to. Further details about the data collection and analysis procedures, as well as on the (live) GPS metadata, can be found in the Methods section. Overall, considering the recent literature, and to the best of our knowledge, the amount of GPS-tagged data we have gathered is certainly remarkable not only in terms of volume, but also for the covered geographical and temporal extension. 

\begin{figure}[!h]
\begin{centering}
\includegraphics[width=1.0\textwidth,angle=0]{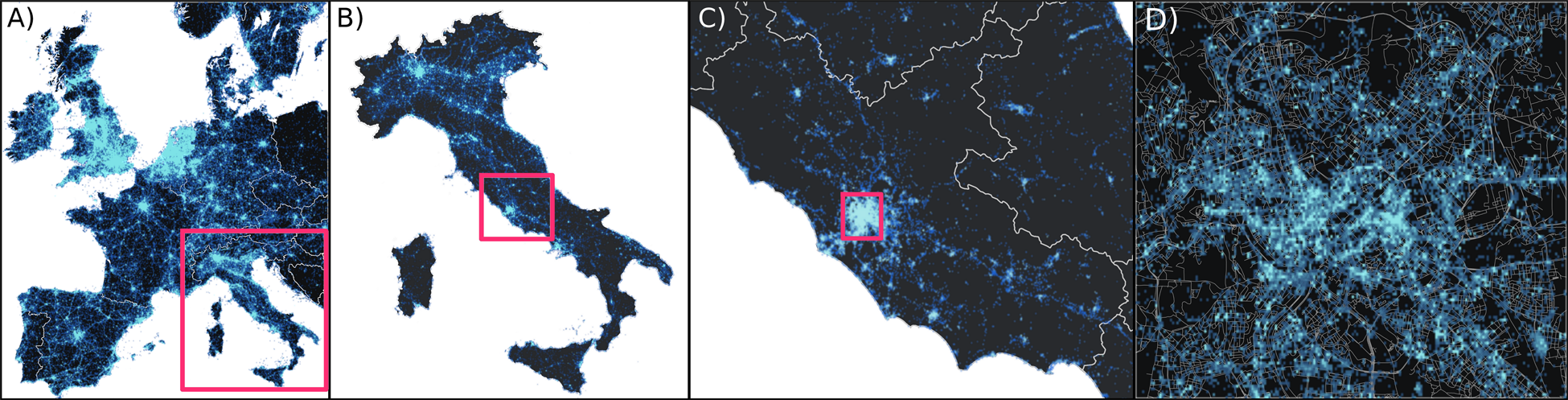}
\caption{\label{fig:multiscale} {\bf Multiscale view of the geolocated Twitter signal.} The large number of geolocated Twitter traffic allows for a high resolution characterization of human behavior. A) Europe B) Italy C) Lazio region D) Rome. The squares highlight the zooming areas..}
\end{centering}
\end{figure}

\begin{figure}[!h]
\begin{centering}
\includegraphics[width=0.6\columnwidth]{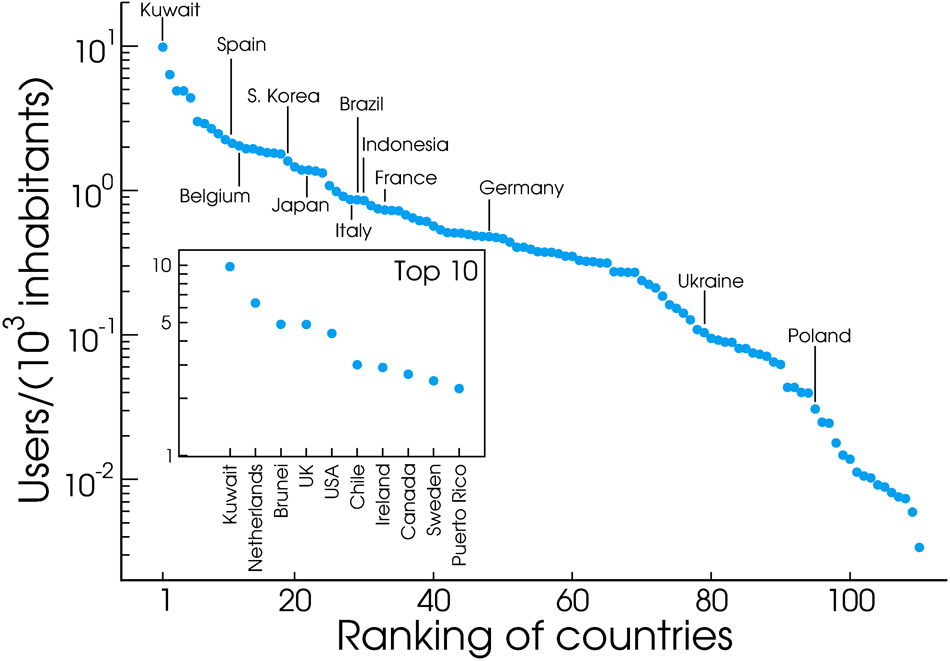}
\caption{\label{fig:userscapita_rank} {\bf Ranking of countries by users per capita.} Ranking of countries as per average number of Twitter users over a population of $1000$ individuals.}
\end{centering}
\end{figure}

\begin{figure}[!h]
\begin{centering}
\includegraphics[width=0.6\columnwidth]{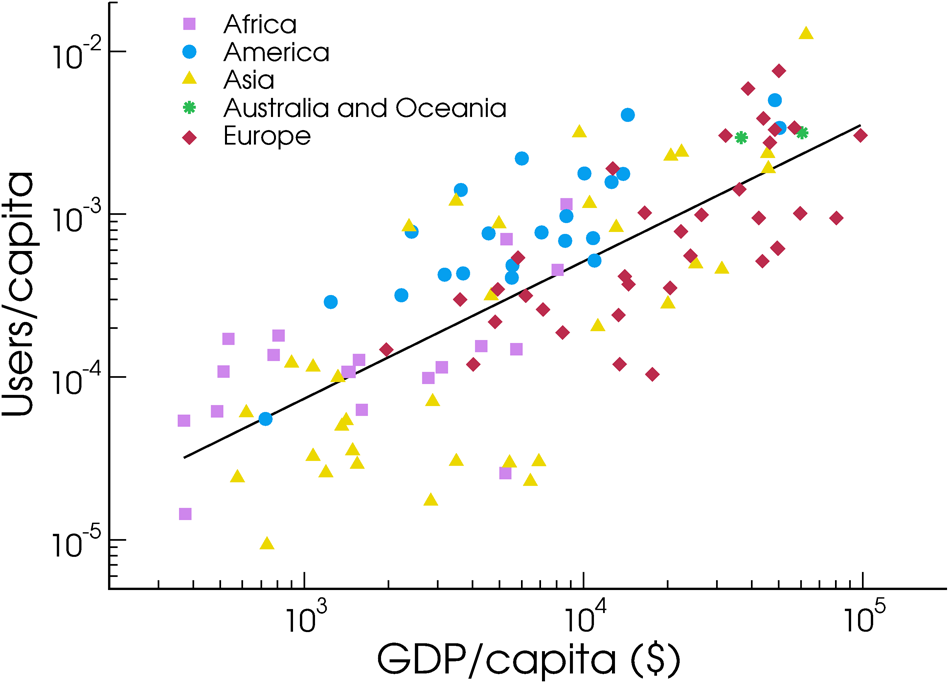}
\caption{\label{fig:gdp} {\bf Users and GDP per capita.} Correlation between country level Twitter penetration and GDP/capita.}
\end{centering}
\end{figure}

Fig.~\ref{fig:multiscale} illustrates the potential of inspection at different resolutions, from continent to city level, highlighting the detailed structure that is visible at each scale. Countries are easily identified along with their major metropolitan areas, and even within specific cities it is possible to observe a high degree of details. Coupling this geographical resolution with language detection tools (see Methods) provides us with a remarkable view of how languages are used in different areas. However, Twitter adoption is not homogeneous across different countries.  Fig.~\ref{fig:userscapita_rank} ranks countries in descending order in terms of Twitter adoption, defined as the ratio between Twitter users and total population (i.e. Twitter users per $1,000$ inhabitants). The emerging picture is highly heterogeneous, as expected, since our data come exclusively from smartphone devices that are consequentially tied to the availability of local infrastructures. In order to support the hypothesis that economic diversity is  a primary source of heterogeneity in the Twitter adoption (in mobile devices), we investigated whether the Gross Domestic Product (GDP) of a country could serve as a predictor of microblogging adoption. Fig.~\ref{fig:gdp} shows that this is the case, the GDP and the Twitter users per capita being clearly correlated.  Moreover, different continents (identified by different color codes in Fig.~\ref{fig:gdp}) cluster together, indicating, that cultural as well as socio-economic factors concur at once in determining the observed pattern.

\begin{figure*}[!h]
\begin{centering}
\includegraphics[width=1.0\textwidth]{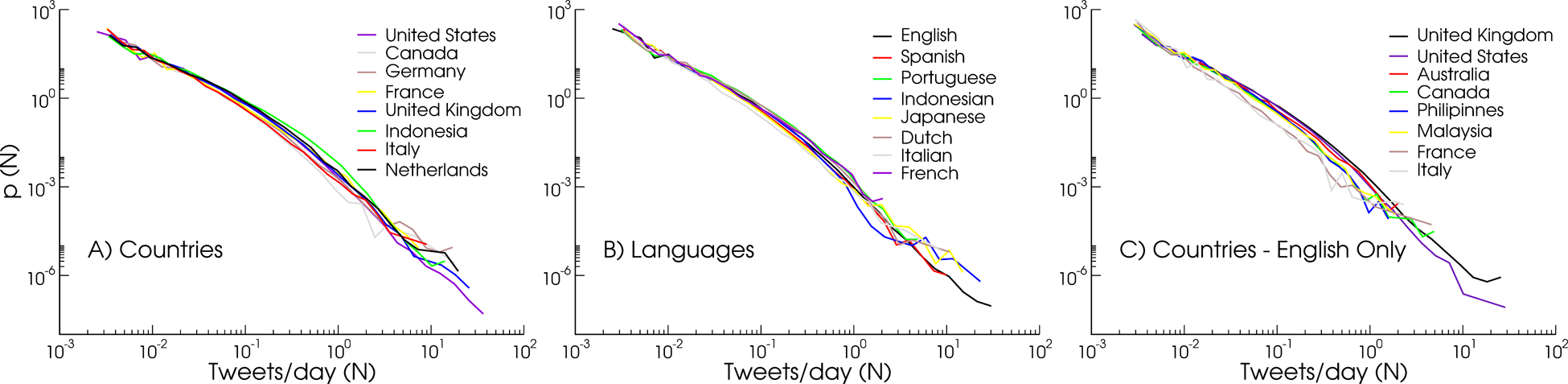}
\caption{\label{fig:useract} {\bf User Activity.} Probability density $p(N)$ of user activity (number of daily tweets N) grouped by country (A) and language (B), and by country while considering English tweets exclusively (C). Different curves collapse naturally, without any functional rescaling, indicating the presence of a seemingly universal distribution of users activity, independent from cultural backgrounds.}
\end{centering}
\end{figure*}

Geographical analyses at any scale require the aggregation of the signal produced by different users, and it is crucial to have a clear understanding of the patterns of single user activity. One might suspect that usage patterns at the individual level may show large heterogeneities across country and thus cultures. In order to test statistically the presence of different usage patterns we gather the number of tweets per unit time sent by each single identified user. From this data we construct the probability density function $p\left(N\right)$ that any given user emits $N$ tweets per considered unit time. In our analysis we considered as reference unit time one day. Furthermore, the $p\left(N\right)$ distribution can be analyzed by restricting the statistical analysis to users belonging to a specific country, a specific language or both. Interestingly, Fig. \ref{fig:useract} shows that the distributions exhibit a universal shape irrespective both of country (panel A),  language (panel B), or the weight of each countries on specific languages (panel C).  As we will see this finding is pivotal for an unbiased comparison of different geographical and linguistic scenarios. Any dependence of the activity distribution upon the language or location of the users would have reduced the array of possible analysis. It is worth stressing also that the curves overlap each other naturally, i.e., with no need for any rescaling or transformation. Although this feature indicates a very strong statistical homogeneity at the population level,  the observed distribution turns out to span almost $4$ orders of magnitude. The broad nature of this universal distribution is clear evidence of strong individual level heterogeneity. For this reason, in order to avoid distortions due to extremely active users, we consider only the proportion of tweets emitted by each user in a given language. Thus, a user $i$ that tweets in a set, $L$, of different languages, $L=\{A,B,C,\ldots, Z \}$, will contribute to each language $X$ for a fraction $N^i_X/\sum_{Y}N^i_Y$. We define $N^i_X$ the total number of tweets written by the user in language $X$. We adopt the same normalization also for the position of the user. The reasons for this normalization are multiple. First, the amount of tweets collected for each user ranges over several orders of magnitude. Very active users, as well as automatic bots, might therefore distort or mask the signal coming from ``common" individuals. Second, tourism might be a strong source of noise when trying to understand the demographics of a country or of a city. Touristic locations in the South of France or Italy might for example exhibit a high proportion of tweets in English or German. 

\subsection{Language analysis at different geographic scales}
\label{sec:language_analysis}

\begin{figure}[!h]
\begin{centering}
\includegraphics[width=0.6\columnwidth]{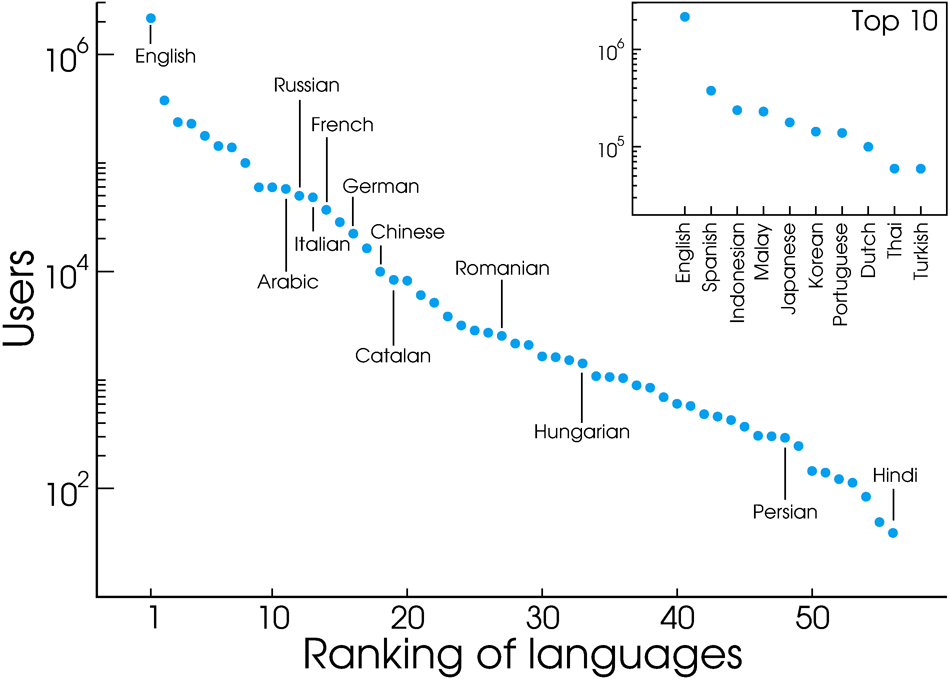}
\caption{\label{fig:lang_global} {\bf Languages by number of users.} Languages ranked by total number of users. For clarity, only languages with more than $30$ users are shown.}
\end{centering}
\end{figure}

\begin{figure}[!h]
\begin{centering}
\includegraphics[width=1.0\textwidth,angle=0]{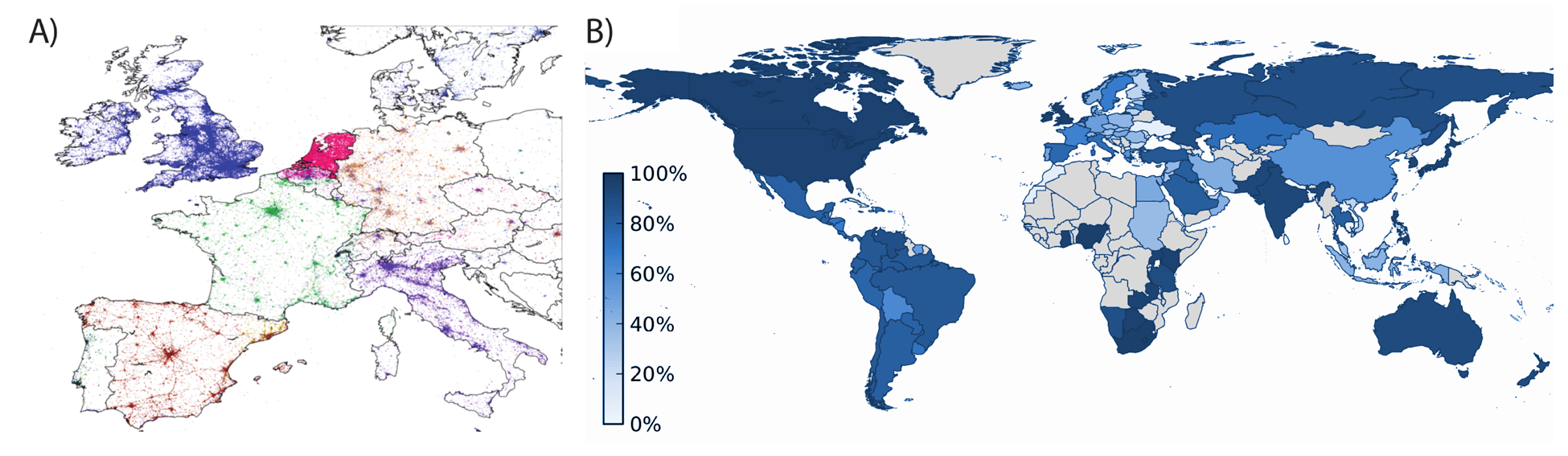}
\caption{\label{fig:country} {\bf Geographic distribution of languages around the world.} A) Raw Twitter signal. Each color corresponds to a language. Densely populated areas are easily identified, while, as expected, languages are well separated among European countries. B) Dominant language usage. The color of each country indicates the fraction of users adopting the official language in tweets. Gray represent countries without statistically significant signal.}
\end{centering}
\end{figure}

The ranking of languages in our signal is presented in Fig.~\ref{fig:lang_global}, where the ordering is determined by the number of users we observe for each one of them. As expected, English is largely dominant. Spanish occupies the second position despite being almost $6$ times less popular. Interestingly, these languages are followed by Malay and Indonesian, reflecting the fact that Indonesia is a very active country in absolute terms, even though in terms of users per capita the country is only ranked in the $30th$ position (see Fig.~\ref{fig:country}). Here the effect of each countries population size becomes clear. A large country as Indonesia does not need a large per capita Twitter penetration to make its language very visible in Twitter, while much smaller Netherlands does. And in fact the Netherlands is the second country in terms of users per capita (see Fig.~\ref{fig:country}), making Dutch the $8th$ most common language.

It is worth stressing that our statistics \textit{do not} reflect the overall estimates of language speakers in the world. According to Ethnologue: Languages of the World \cite{ethnologue},  when native and secondary speakers are considered together Standard Chinese leads the ranking ($1.0\times10^9$ speakers), followed by English ($5.0\times10^8$ speakers), Spanish ($3.9\times10^8$ speakers), Hindi ($3.0\times10^8$ speakers) and Russian ($2.5\times10^8$ speakers), with Malay/Indonesian ranked as $8th$ ($1.6\times10^8$ speakers). These discrepancies do not prevent us from extracting meaningful information in countries where Twitter is sufficiently high to serve as an accurate mirror of the population, but it serves as a reminder that we
are observing the worldwide linguistic landscape through the lenses of a (specific) microblogging platform which, for example, is not available in China. Also the age composition of Twitter users must be taken into account if one is to compensate for differences with respect to the official census data~\cite{mislove11-1}.

\begin{figure}[!h]
\begin{centering}
\includegraphics[width=0.5\columnwidth,angle=0]{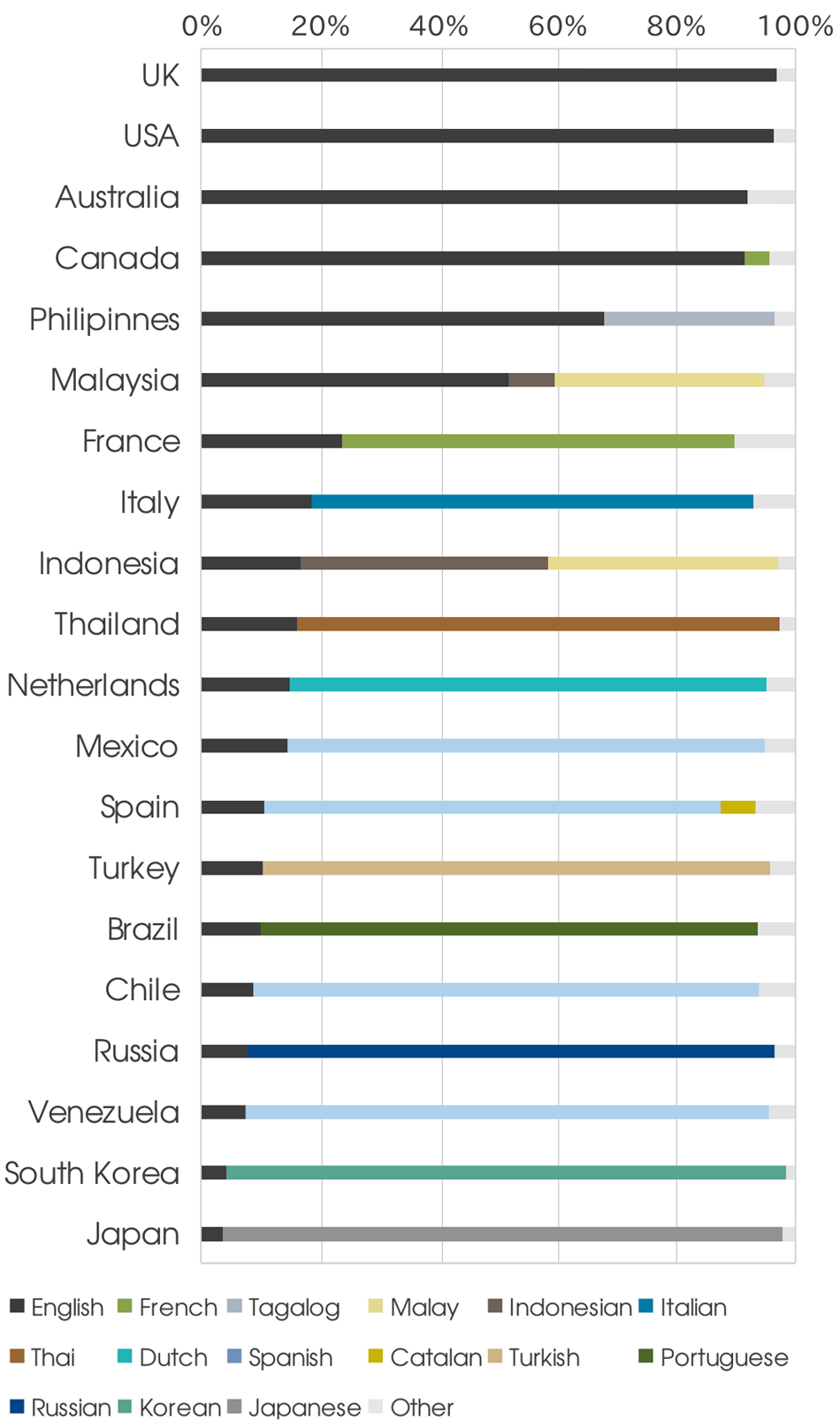}
\caption{\label{fig:top20} {\bf Language share of the most active countries.} Language adopted by users coming from Top $20$ most active countries, ordered by number of English tweets.}
\end{centering}
\end{figure}

\emph{Country level.}  When we color each tweet according to its language and display them on a map we see immediately that most content produced within each country is written in its own dominant language (see Fig~\ref{fig:country}-A). This is further confirmed in Fig~\ref{fig:country}-B, which shows the extent to which the dominant language prevails over other idioms in each country. In Figure~\ref{fig:top20} we plot, for each of the Top $20$ countries (by number of tweets), the fraction of users tweeting in each language.  Interestingly, countries like France and Italy, which are characterized by a well defined and substantially homogeneous linguistic identity, emit more than $20 \%$ of their tweets in English and other languages. Since the most common language in Twitter is English, this is perhaps not surprising. It is in fact reasonable that even users of non-English speaking countries choose to Tweet in English as a form of reaching out to a broader audience. 

\begin{figure}[!h]
\begin{centering}
\includegraphics[width=0.7\columnwidth]{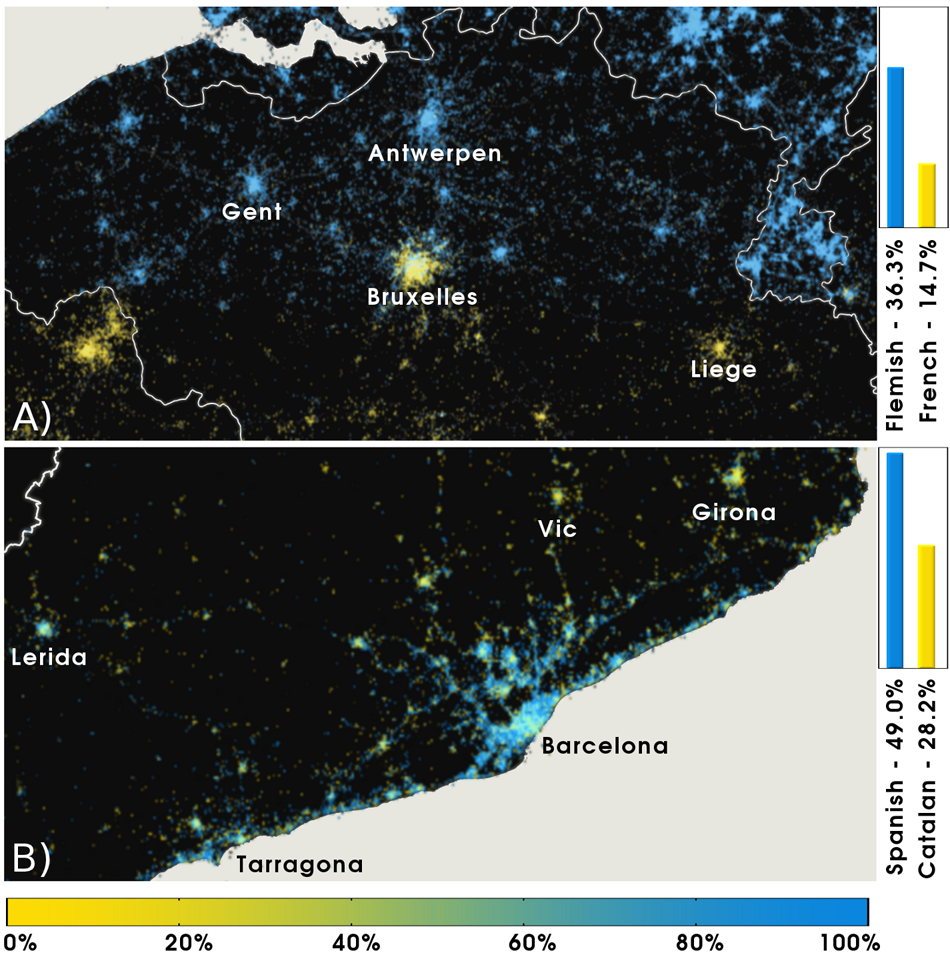}
\caption{\label{fig:regions} {\bf Language polarization in Belgium and Catalonia, Spain.} In each cell ($600 m$ resolution) we compute the user-normalized ratio between the two languages being considered in each case.  A) Belgium. B) Catalonia. The color bar is labeled according to the relative dominance of the language denoted by blue.}
\end{centering}
\end{figure}

\emph{Regional level.} To understand the geospatial heterogeneity of different linguistic backgrounds, we drill down data to  small - within-country- scales.
It is interesting, for instance, to look at the spatial distribution of the different languages in multilingual regions. Figure \ref{fig:regions}-A illustrates the geographical distribution of languages used in Belgium, where the North part of the country uses predominantly Flemish, while in the South of the country the dominant language is (Walloon) French. Overall, Flemish accounts for $36.3\%$ of the users, while French is the language of $14.7\%$ of the users within the country borders, i.e. Dutch is $2.5$ times more popular than French. Census data set the Dutch to French ratio (as first Languages) to $1.5$ \cite{stat_belgium}. The result emerging from the Twitter analysis is qualitatively correct, the quantitative mismatch being  explained by the different Twitter penetration in neigboring France and Netherlands, whose dominant language is of course French and Dutch. In the first case, the number of users per $1000$ inhabitants is $0.85$, while in the second is $6.34$, more than $7$ times higher (see also Fig.~\ref{fig:userscapita_rank}). The Dutch speaking population of Belgium finds itself embedded in a much richer Twitter environment, and consequently is more involved in the microblogging activity.

Moving to a within-country scale, Figure \ref{fig:regions}-B shows the linguistic distribution in Catalonia, an autonomous region of Spain. Here Catalan and Spanish are clearly intermixed (particularly in Barcelona), even though Spanish is the most popular language, with a share of $49.0 \%$ of the users where Catalan represents $28.2 \%$ of the signal, making that Spanish $1.7$ times more popular than Catalan. Interestingly, the Spanish to Catalan ratio is $1.25$ when the habitual language of adults living in Catalonia is considered, according to a survey performed in $2008$ by the Institute of Statistics of Catalonia  \cite{catalonia}. In this case the Twitter data is close to the census data, although some considerations are in order. First, census data do not take into account the presence of tourists, whose Twitter activity is on the other hand recorded. Second, Twitter users may be biased towards the most common languages, in order to reach a wider audience. This interpretation is corroborated by the fact that while in our dataset Catalan and Spanish account for the $77.2\%$ of the users, they represent the habitual language of $93.5\%$ of the population according to the above mentioned survey. In the same way, English, which according to census data is customarily spoken by less than $0.01\%$ of the resident population, is adopted by $15.2\%$ of the users. Going at a deeper level of inspection, we see that the  Catalan language is more widely used in the central and Northern part of the region than in the area of Barcelona and the coast connecting this city to Tarragona. Remarkably, this pattern agrees with the overall picture provided by census data \cite{catalonia}, thus confirming once again the validity of online data in providing meaningful informations, even at the within-country scale.

\begin{figure}[!h]
\begin{centering}
\includegraphics[width=0.7\columnwidth]{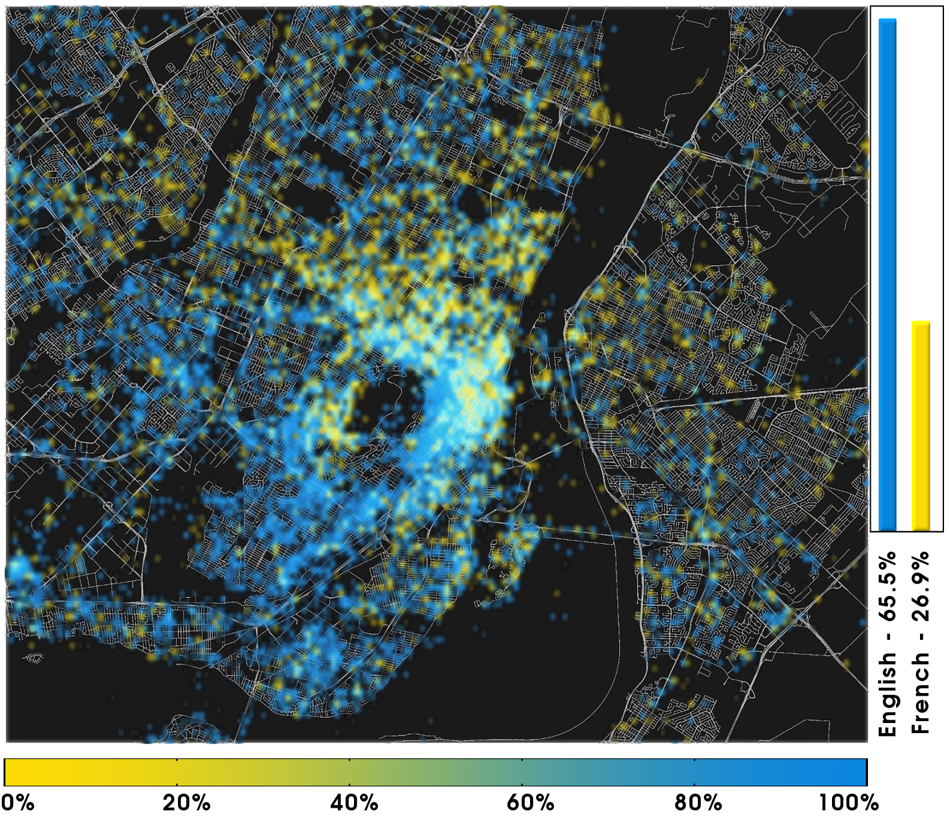}
\caption{\label{fig:montreal}  {\bf Language polarization in Montreal, QC, Canada.} English and French are considered. In each cell ($200 m \times 200 m$) we compute the user-normalized ratio between English and French (excluding all other languages). Blue - English, Yellow - French. The color bar is labeled according to the relative dominance of English to French.}
\end{centering}
\end{figure}

\emph{City level.} The high quality of the GPS geolocalized signal allows the inspection of the language demographics of single cities. Figure \ref{fig:montreal} shows the city of Montreal, where English and French are the most used languages. While English is significantly more popular ($65.5 \%$ of users, vs. the French $26.9 \%$), there appear to be spatial segregation, with French being more popular in the northern neighborhoods. Overall, the English is $2.4$ times more popular than French in our signal, while the situation is the opposite according to census data surveying languages spoken at home, where French is $3.1$ times more frequent than English \cite{stat_canada}. This reversal is not easy to interpret, but we speculate that the geographical location of Montreal, and the fact that we do not consider the entire metropolitan population, along with the fact that English is in general the privileged communication language in North America, are two factors that might play an important role.

\begin{figure}[!h]
\begin{centering}
\includegraphics[width=0.7\columnwidth]{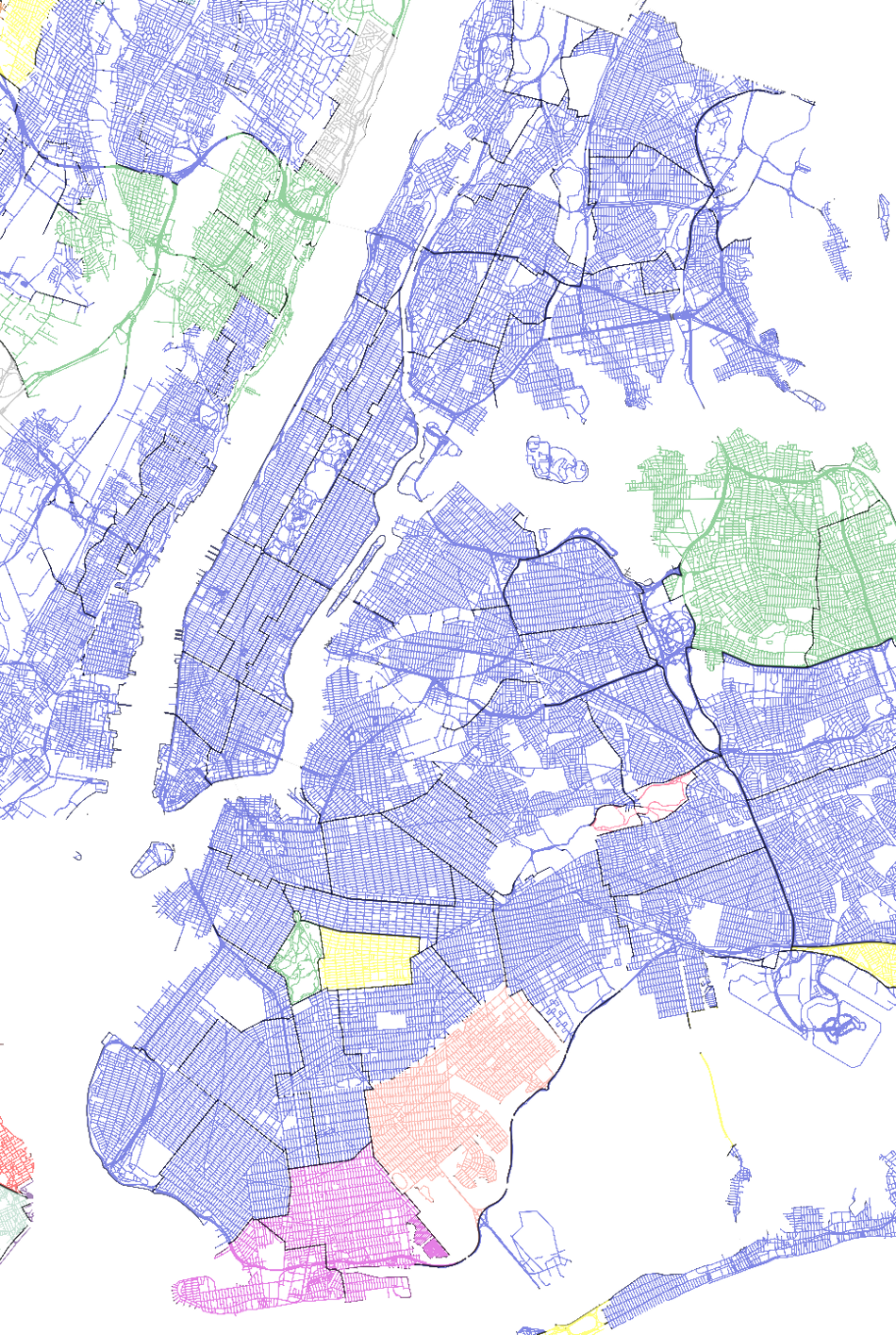}
\caption{\label{fig:NYC} {\bf Language polarization in New York City, NY, USA.} The second language by district or municipality (in the case of New Jersey state) is shown. Blue - Spanish, Light Green - Korean, Fuchsia - Russian, Red - Portuguese, Yellow - Japanese, Pink - Dutch, Grey - Danish, Coral - Indonesian}
\end{centering}
\end{figure}

The same analysis can be performed at the level of city neighborhood. In the case of New York City, a city known for its cultural diversity, several non-English speaking communities are already well-defined and documented \cite{lobo2002impact,nyc_kor1,nyc_kor2,nyc_dutch,nyc_russian}. For this case study, we partition NYC, Long Island, and New Jersey state into districts, towns, and municipalities, respectively. We do not consider the signal in English (since it is the official language, and homogeneously predominant in the area) and we focus instead on the language exhibiting the second largest number of users inside each district/town. Some of the most popular communities are those of Spanish speakers in Harlem, Bronx, and parts of Queens \cite{lobo2002impact}. However, Spanish is shared by people from many different cultural backgrounds and it is also widely used across the United States. It is thus difficult to estimate the exact location and dimensions of these communities solely based on Twitter signal. In fact, it is clear that Spanish dominates as a second language in a number of districts of Figure \ref{fig:NYC}. Remarkable, on the other hand, is the clear delimitation of other communities. The Korean communities in Palisades Park, NJ and Flushing, NY are of considerable size and also very socially active \cite{nyc_kor1,nyc_kor2}. Marine Park, NY, on the other hand, has a long history of Dutch immigration that dates back to the first European settlers in the area \cite{nyc_dutch}. Another notable example is the case of Coney Island, NY, which is home to the largest Russian community in the United States \cite{nyc_russian}.The high resolution of our dataset  allows us to visualize these communities without any a priori assumptions.

\subsection{Seasonal variations}
\label{sec:seasonal}

\begin{figure}[!h]
\begin{centering}
\includegraphics[width=1.0\textwidth]{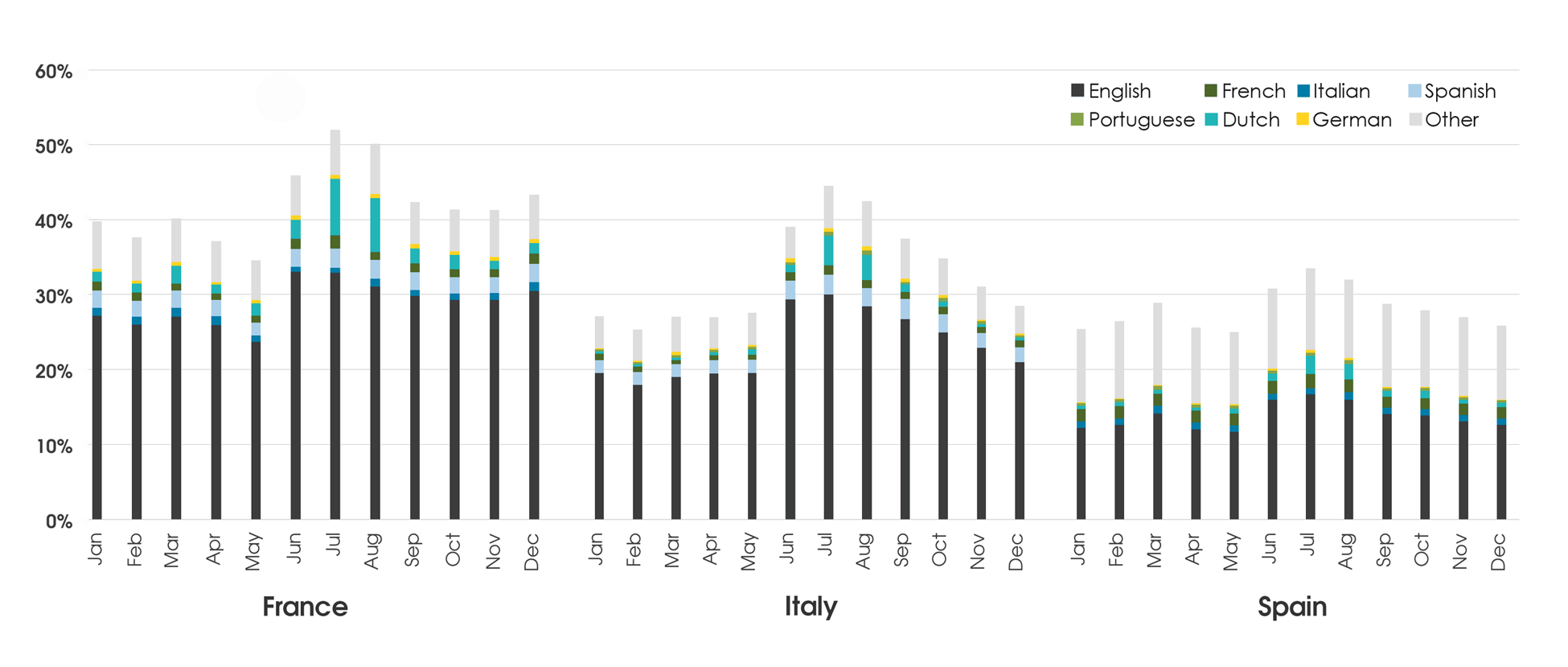}
\caption{\label{seasonal} {\bf Monthly variations in Language use.} Fraction of minority languages in specific countries as a function of the month. Increases in a specific language share indicate the presence of tourists visiting the country. Peaks are clearly visible during the local summer period.}
\end{centering}
\end{figure}

Now that we have a good characterization of the relative linguistic composition of each country we can assess the of use our data to study and analyze seasonal variations of language composition, as this would give us valuable insights onto population movements occurring over the course of a year. In particular, we might expect that during more touristic seasons one could observe a relative decrease in traffic occurring in the local dominant language and a corresponding increase in content being generated in foreign languages. In Fig.~\ref{seasonal} we show the relative contributions of minority languages from users within a given country as a function of the month of the year.  In particular we single out traditional touristic destinations, such as France, Italy, and Spain, where clear variations are indeed visible during the summer. 

Our analysis allows not only to identify the aggregate touristic fluxes, but also to infer the regions of origin on the basis of the observed language. Of course, the pattern we observe are certainly slightly biased by the specificity of our observation point, so that for example the contribution of Dutch is likely to be constantly overestimated due to the high penetration of Twitter in the Netherlands. However, the possibility of observing seasonal fluxes is absolutely remarkable if we consider the low cost, both in terms of time and resources, that a Twitter survey requires, compared to more traditional approaches. Moreover, monitoring social networks allows us to gain a real-time perspective of the fluxes, which is of course extremely hard to achieve through demographic studies. 

\section{Discussion}
\label{sec:discussion}

In this paper we have characterized the worldwide linguistic geography as observed from the Twitter platform, aggregating microblogging data at different scales, from country level down to the neighborhood scale. Although we show that Twitter penetration is highly heterogeneous and closely correlated with GDP, we find that the statistical usage pattern of the microblogging platform turns out to be independent from such factors as country and language. This feature allows us to address different issues, such as linguistic homogeneity at the country level, the geographic distribution of different languages in bilingual regions or cities, and the identification of linguistically specific urban communities. Focusing on specific case-studies, we have shown that while Twitter trends mirror census data quite accurately, even though specific deviations might emerge when comparing data that can be influenced by the adoption rate of the microblogging platform or the fact that English is the most widely used language in Twitter. Finally, the analysis of temporal variations of the language composition of a given country opens up the possibility of observing traveling patterns and identifying in real time seasonal traveling and mobility patterns. The presented results confirms the potential and opportunities offered by open access data -such as microblogging posts- in the characterization and analysis of demographic and social phenomena.

\section{Materials and Methods}
\label{sec:methods}

\subsection*{Data Collection}\label{ssec:data} The datased was obtained by extracting tweets from the raw Twitter Gardenhose feed~\cite{ratkiewicz11-2}. The Gardenhose is an unbiased sample of $10\%$ of the entire number of tweets  provides a statistically significant real time view of all activity within the Twitter ecosystem. Twitter added support for explicit geotagging of tweets since November $2009$, by providing API hooks that could be used by third party developers to embedded GPS coordinates within the metadata of each tweet. Since high quality GPS systems are increasingly common in mobile devices, this feature immediately became popular with mobile application developers and is currently available in hundreds of different twitter clients. On average, about $1\%$ of the tweets contain GPS information

\subsection{Language Detection} Automatically determining the language in which a certain text was written is problem of great practical importance for machine learning and data mining. Perhaps the better known example of this is a feature in Google's popular web browser, Chrome, that offers to translate a page from it's original language to the users native language has a feature that offers to translate a page to the users preferred language. The library that detects the original language of the page leverages Googles extensive experience with data mining and has been extracted from Chromes source code and made available separately as the ``Chromium Compact Language Detector''~\cite{chromium}, a library that was extracted from the open source version of Google's Chrome browser that is currently in use by millions of browsers around the world. To further ensure the accuracy of the result, we filter the results by using an uncertainty threshold within the language detector.

\subsection{Geolocalization and Statistics}

We restrict our analysis to tweets containing GPS coordinates, i.e. generated by using a smartphone with an Internet connection. This choice allows for the maximum geographical resolution, but inevitably reduces the volume of available signal. In fact, the data we have used for this paper constitutes just about $1 \%$ of the signal we have collected, which on its turn is approximately $10\%$ of the total Twitter volume.

The amount of geolocalized tweets could be increased by considering self-reported informations. In fact, users are encouraged to provide their location information in the user profile, but it is not subject to any format restriction. Moreover, Twitter platforms do not prompt the user for an update of this field, thus any change to this metadata field has to be spontaneous and made voluntarily. For this reason, the information in the user profile is sometimes erroneous or has low granularity. While the research community is on a continuous quest to understand how to mine and geocode this data, doing so brings about many challenges \cite{justin_bieber}. Moreover, when addressing temporal variations in mobility patterns, the use of smartphone GPS coordinates is required. 

The metadata accompanying a tweet may also contain the geographical coordinates of a previous location in the field of self-reported location. These `historical' locations might bias statistical measures involving mobility and/or fine graining, thus we considered them only in generating the language maps (Belgium, Catalonia, NYC). 
All sets of analysis performed at the country level make use solely of live-GPS coordinates. We consider only those countries for which our signal is generated by at least $200$ users, normalized by their activity and location. So if a user emits $30\%$ of her tweets from a given country she will contribute as $0.3$ users to that country.  $110$ countries satisfy this minimum user threshold.

Finally, it is crucial stressing that every set of statistical measures performed in this paper is done at the user level, in order to reduce the noise that bots or cyborgs might add to the analysis. If not suitably addressed, in fact, their presence could induce wrong conclusions on the day-to-day behavior of the average person \cite{chu_2010}.

\section{Acknowledgments} We acknowledge the support by the NSF ICES award CCF-1101743. For the
analysis of data data outside of the USA we acknowledge the Intelligence Advanced Research Projects Activity (IARPA) via
Department of Interior National Business Center (DoI / NBC) contract
number D12PC00285.  The views and conclusions contained herein are
those of the authors and
should not be interpreted as necessarily representing the official
policies or endorsements, either expressed or implied, of IARPA,
DoI/NBE, or the U.S. Government.


\begin{thebibliography}{10}
\providecommand{\url}[1]{\texttt{#1}}
\providecommand{\urlprefix}{URL }
\expandafter\ifx\csname urlstyle\endcsname\relax
  \providecommand{\doi}[1]{doi:\discretionary{}{}{}#1}\else
  \providecommand{\doi}{doi:\discretionary{}{}{}\begingroup
  \urlstyle{rm}\Url}\fi
\providecommand{\bibAnnoteFile}[1]{%
  \IfFileExists{#1}{\begin{quotation}\noindent\textsc{Key:} #1\\
  \textsc{Annotation:}\ \input{#1}\end{quotation}}{}}
\providecommand{\bibAnnote}[2]{%
  \begin{quotation}\noindent\textsc{Key:} #1\\
  \textsc{Annotation:}\ #2\end{quotation}}
\providecommand{\eprint}[2][]{\url{#2}}

\bibitem{gonzalez08-1}
Gonz{\'a}lez MC, Hidalgo CA, Barab{\'a}si AL (2008) Understanding individual
  human mobility patterns.
\newblock Nature 453: 779.
\input{gonzalez08-1}

\bibitem{Onnela:2007p4919}
Onnela JP, Saramaki J, Hyvonen J, Szabo G, Lazer D, et~al. (2007) {Structure
  and tie strengths in mobile communication networks}.
\newblock Proceedings of the National Academy of Sciences 104: 7332--7336.
\input{Onnela:2007p4919}

\bibitem{hale2012world}
Hale S, Gaffney D, Graham M (2012) Where in the world are you? geolocation and
  language identification in twitter.
\newblock Technical report.
\input{hale2012world}

\bibitem{conover11-2}
Conover M, Ratkiewicz J, Gon\c{c}alves B, Haff J, Flammini A, et~al. (2011)
  Predicting the political alignment of twitter users.
\newblock In: IEEE Third International Conference on Social Computing
  (SOCIALCOM). p. 192.
\input{conover11-2}

\bibitem{sang2012predicting}
Sang E, Bos J (2012) Predicting the 2011 dutch senate election results with
  twitter.
\newblock EACL 2012 : 53.
\input{sang2012predicting}

\bibitem{goncalves2011modeling}
Gon{\c{c}}alves B, Perra N, Vespignani A (2011) Modeling users' activity on
  twitter networks: Validation of dunbar's number.
\newblock PLoS One 6: e22656.
\input{goncalves2011modeling}

\bibitem{borge2011structural}
Borge-Holthoefer J, Rivero A, Garc{\'\i}a I, Cauh{\'e} E, Ferrer A, et~al.
  (2011) Structural and dynamical patterns on online social networks: the
  spanish may 15th movement as a case study.
\newblock PLoS One 6: e23883.
\input{borge2011structural}

\bibitem{tumasjan2010predicting}
Tumasjan A, Sprenger T, Sandner P, Welpe I (2010) Predicting elections with
  twitter: What 140 characters reveal about political sentiment.
\newblock In: Proceedings of the Fourth International AAAI Conference on
  Weblogs and Social Media. pp. 178--185.
\input{tumasjan2010predicting}

\bibitem{culotta2010towards}
Culotta A (2010) Towards detecting influenza epidemics by analyzing twitter
  messages.
\newblock In: Proceedings of the First Workshop on Social Media Analytics. ACM,
  pp. 115--122.
\input{culotta2010towards}

\bibitem{salathe11-1}
Salathe M, Khandelwal S (2011) {Assessing Vaccination Sentiments with Online
  Social Media: Implications for Infectious Disease Dynamics and Control}.
\newblock PLoS Computational Biology 7: e1002199.
\input{salathe11-1}

\bibitem{salathe12-2}
Salathe M, Bengtsson L, Bodnar TJ, Brewer DD, Brownstein JS, et~al. (2012)
  {Digital Epidemiology}.
\newblock PLoS Comput Biol 8: E1002616.
\input{salathe12-2}

\bibitem{kulshrestha2012geographic}
Kulshrestha J, Kooti F, Nikravesh A, Gummadi K (2012) Geographic dissection of
  the twitter network.
\newblock In: In Proceedings of the 6th International AAAI Conference on
  Weblogs and Social Media (ICWSM).
\input{kulshrestha2012geographic}

\bibitem{mislove2011understanding}
Mislove A, Lehmann S, Ahn Y, Onnela J, Rosenquist J (2011) Understanding the
  demographics of twitter users.
\newblock In: Fifth International AAAI Conference on Weblogs and Social Media.
\input{mislove2011understanding}

\bibitem{hong_2011}
Hong L, Convertino G, Chi E (2011) Language matters in twitter: A large scale
  study.
\newblock In: International AAAI Conference on Weblogs and Social Media. pp.
  518--521.
\input{hong_2011}

\bibitem{Giannotti2012}
Giannotti F, Pedreschi D, Pentland A, Lukowicz P, Kossmann D, et~al. (2012) A
  planetary nervous system for social mining and collective awareness.
\newblock The European Physical Journal Special Topics 214: 49-75.
\input{Giannotti2012}

\bibitem{williams88-1}
Williams CH, editor (1988) Language in Geographic Context.
\newblock Multilingual Matters, Ltd.
\input{williams88-1}

\bibitem{baronchelli2012language}
Baronchelli A, Loreto V, Tria F (2012) Language dynamics.
\newblock Advances in Complex Systems 15, 1203002.
\input{baronchelli2012language}

\bibitem{poblete2011all}
Poblete B, Garcia R, Mendoza M, Jaimes A (2011) Do all birds tweet the same?:
  characterizing twitter around the world.
\newblock In: Proceedings of the 20th ACM international conference on
  Information and knowledge management. ACM, pp. 1025--1030.
\input{poblete2011all}

\bibitem{Weerkamp2011}
Weerkamp W, Carter S, Tsagkias M (2011) How people use twitter in different
  languages.
\newblock In: Proceedings of the ACM WebSci'11, June 14-17 2011, Koblenz,
  Germany. p.~1.
\input{Weerkamp2011}

\bibitem{takhteyev2012geography}
Takhteyev Y, Gruzd A, Wellman B (2012) Geography of twitter networks.
\newblock Social Networks 34: 73--81.
\input{takhteyev2012geography}

\bibitem{ethnologue}
 (Retrieved Dec. 2012).
\newblock {Languages of the world. Summary by language size}.
\newblock
  \urlprefix\url{http://www.ethnologue.org/ethno_docs/distribution.asp?by=size%
}.
\input{ethnologue}

\bibitem{mislove11-1}
Mislove A, Lehmann S, Ahn YY, Onnela JP, Rosenquist JN (2011) Understanding the
  demographics of twitter users.
\newblock In: In Proceedings of the Fifth International AAAI Conference on
  Weblogs and Social Media.
\input{mislove11-1}

\bibitem{stat_belgium}
 (Retrieved Dec. 2012).
\newblock Europeans and their languages.
\newblock
  \urlprefix\url{http://ec.europa.eu/public_opinion/archives/ebs/ebs_243_en.pd%
f}.
\input{stat_belgium}

\bibitem{catalonia}
 (Retrieved Sept. 2012).
\newblock Usos ling\"{u}\'istics. llengua inicial, d'identificaci\'o i
  habitual.
\newblock \urlprefix\url{http://www.idescat.cat/dequavi/?TC=444&V0=15&V1=2}.
\input{catalonia}

\bibitem{stat_canada}
 (Retrieved Dec. 2012).
\newblock Population by language spoken most often at home and age groups, 2006
  counts, for canada, provinces and territories, and census subdivisions
  (municipalities) with $5,000$-plus population - $20\%$ sample data.
\newblock
  \urlprefix\url{http://www12.statcan.ca/census-recensement/2006/dp-pd/hlt/97-%
555/T402-eng.cfm?Lang=E&T=402&GH=7&GF=24&G5=1&SC=1&RPP=100&SR=1&S=1&O=D&D1=1}.
\input{stat_canada}

\bibitem{lobo2002impact}
Lobo A, Flores R, Salvo J (2002) The impact of hispanic growth on the
  racial/ethnic composition of new york city neighborhoods.
\newblock Urban Affairs Review 37: 703--727.
\input{lobo2002impact}

\bibitem{nyc_kor1}
 (2012).
\newblock \url{http://njmonthly.com/articles/best-of-Jersey/seoul_mates.html}.
\input{nyc_kor1}

\bibitem{nyc_kor2}
 (2012).
\newblock \url{http://www.kcsny.org/}.
\input{nyc_kor2}

\bibitem{nyc_dutch}
 (2012).
\newblock \url{https://www.nycgovparks.org/parks/marinepark/history}.
\input{nyc_dutch}

\bibitem{nyc_russian}
 (2012).
\newblock
  \url{http://offmetro.com/ny/2008/04/13/brighton-beach-a-voyage-to-russia/}.
\input{nyc_russian}

\bibitem{ratkiewicz11-2}
Ratkiewicz J, Conover M, Meiss M, Gon\c{c}alves B, Patil S, et~al. (2011)
  Truthy: Mapping the spread of astroturf in microblog streams.
\newblock Twentieth International World Wide Web Conference 249.
\input{ratkiewicz11-2}

\bibitem{chromium}
Candless MM (2012).
\newblock http://code.google.com/p/chromium-compact-language-detector/.
\input{chromium}

\bibitem{justin_bieber}
Hecht B, Hong L, Suh B, Chi EH (2011) Tweets from justin bieber's heart: the
  dynamics of the location field in user profiles.
\newblock In: Proceedings of the SIGCHI Conference on Human Factors in
  Computing Systems. New York, NY, USA: ACM, CHI '11, pp. 237--246.
\newblock \doi{10.1145/1978942.1978976}.
\newblock \urlprefix\url{http://doi.acm.org/10.1145/1978942.1978976}.
\input{justin_bieber}

\bibitem{chu_2010}
Chu Z, Gianvecchio S, Wang H, Jajodia S (2010) Who is tweeting on twitter:
  human, bot, or cyborg?
\newblock In: Proceedings of the 26th Annual Computer Security Applications
  Conference. New York, NY, USA: ACM, ACSAC '10, pp. 21--30.
\newblock \doi{10.1145/1920261.1920265}.
\newblock \urlprefix\url{http://doi.acm.org/10.1145/1920261.1920265}.
\input{chu_2010}

\end{thebibliography}

\clearpage

\section{Figures}
\vspace{1cm}

\vspace{2cm}


\clearpage

\section{Tables}

\begin{table}[!h]
    \begin{tabular}{l|c}
    
    Days of data collection  & $ 564 $  \\ \hline 
    Tweets/day GPS (live-GPS) & $ 651,400 $ ($128,385$) \\ \hline 
    Users (users live-GPS)  & $ 5,962,976 $ ($4,551,384$) \\ \hline 
    Countries (total) & $ 191 $ \\ \hline 
    Countries (analyzed)  & $ 110 $ \\ \hline 
    Detected languages  & $ 78 $ \\ 

    \end{tabular}
  \caption{ Basic metrics of the data set. Along with the total GPS signal, the fraction of live updates is reported (see Methods for details).} 
  \label{tab:geoloc}
\end{table}

\newpage

\end{document}